\documentclass[aps,pra,twocolumn,superscriptaddress,showkeys]{revtex4}
\usepackage{graphicx}
\usepackage{dcolumn}
\usepackage{color}
\usepackage{bm}
\usepackage{amssymb}
\usepackage{amsmath}
\begin{document}

\title{Observation of dual electromagnetically induced transparencies in cold $^{87}Rb$ atoms}
\author{Charu Mishra}
\email{charumishra@rrcat.gov.in}
\affiliation{Laser Physics Applications Section, Raja Ramanna Center for Advanced Technology,Indore 452013, India}
\affiliation{Homi Bhabha National Institute, Mumbai-400094, India}
\author{A. Chakraborty}
\affiliation{Laser Physics Applications Section, Raja Ramanna Center for Advanced Technology,Indore 452013, India}
\affiliation{Homi Bhabha National Institute, Mumbai-400094, India}
\author{S. P. Ram}
\affiliation{Laser Physics Applications Section, Raja Ramanna Center for Advanced Technology,Indore 452013, India}
\author{S. Singh}
\affiliation{Laser Physics Applications Section, Raja Ramanna Center for Advanced Technology,Indore 452013, India}
\author{V. B. Tiwari}
\affiliation{Laser Physics Applications Section, Raja Ramanna Center for Advanced Technology,Indore 452013, India}
\affiliation{Homi Bhabha National Institute, Mumbai-400094, India}
\author{S. R. Mishra}
\affiliation{Laser Physics Applications Section, Raja Ramanna Center for Advanced Technology,Indore 452013, India}
\affiliation{Homi Bhabha National Institute, Mumbai-400094, India}

\begin{abstract}
We have observed two dual electromagnetically induced transparency (EIT) spectra in probe transmission through cold $^{87}Rb$ atoms trapped in a magneto-optical trap (MOT). These dual EIT peaks are generated in two different N-systems formed by probe, coupling and control beams interacting with different levels in $D_{2}$ line transitions of $^{87}Rb$ atom. The dependence of spectral features of two N-systems on driving beams (\textit{i.e.} coupling and control beams) strength and detuning has been investigated. The results show that the dual EIT spectral features in both the N-systems can be manipulated by varying the strength and detuning of both the driving beams. The two N-systems have been modelled theoretically using a density matrix formalism and results have shown a reasonable agreement with the experimental observations.
\end{abstract}


\maketitle

\section{Introduction}
The interaction of two coherent electromagnetic fields with a three-level atomic system gives rise to an interesting phenomenon known as Electromagnetically Induced Transparency (EIT). EIT is an effect of quantum interference that modifies the optical property of an atomic medium for a weak probe field in presence of another strong coupling field \cite{Harris:1st}. The fascinating applications of EIT includes slow light propagation \cite{Kash:1999, Chen:2009, Alotaibi:2015}, optical switching \cite{Clarke:40:2001, Rao:2017}, precise atomic clocks \cite{Guidry:2017}, tight laser frequency locking \cite{Bell:2007, charu:382:2018}, precise measurement of optical frequency \cite{Iftiquar:281:2008}, non-linear optical process \cite{Doai:2015, Liang:2017}, etc. The basic atomic systems that can exhibit EIT effect are $\Lambda$-\cite{Li:1995, Yan:2001,VBT:2010, Mishina:2011}, ladder-\cite{MOSELEY:1995, Kumar:2009, khoa:2016} and vee-system \cite{Zhao:2002, Kang:2014}. 

The ideal three-level EIT system creates a dark state which does not interact with the probe field and result in the transmission of probe. The researchers have further investigated this dark state by applying pertubation through a microwave field \cite{Wang:2012}, radio-frequency field \cite{Carl:2015} and an additional coupling field \cite{Paspalakis:2002, Qi:2009, Liu:2012, Sabir:2016, Charu:52:2019}, and in these cases a dual EIT structure has been observed. Niu \cite{Niu:2005} had earlier reported that the four-level EIT (dual structure) shows greatly enhanced non-linear effects than an ideal EIT obtained through three-level system. One of the methods to obtain a dual EIT is through the formation of N-system. The first proposal of quantum interference effect in a four-level N-system was given by Harris and Yamamoto \cite{Harris:1998}. Later on, experimental implementation of N-system was carried out by Braje \cite{Braje:2003} and Kang \cite{Kang:2004}. In literature, enhancement in non-linear effects in N-system has also been reported \cite{Schmidt:1996, Kang:2003, Yang:2015, Osman:2017, Kazem:2017}.

In this work, we have implemented an experimental technique to attain two simultaneous N-systems in $D_2$ line transition of cold $^{87}Rb$ atoms which has resulted in two dual EIT peaks in the probe transmission spectrum. The experiments have been performed on cold $^{87}Rb$ atoms trapped in a magneto-optical trap (MOT). In order to realize two N-systems, two driving beams with fixed frequency and one probe beam in scan mode was used in our experiments. The driving beams strength and detuning dependent probe transmission spectrum has been investigated in this work. The results show that the dual EIT spectral features in both the N-systems can be controlled by varying the driving beams strength and detuning. The two N-systems have also been modelled theoretically using a density matrix formalism. The theoretical results explain adequately the experimental observations. The two dual EIT peaks (\textit{i.e.} four EIT peaks) observed experimentally can be useful for application in multi-channel optical communication, slow light propagation, optical switching devices, etc.

\section{Experimental Setup}
\label{sec:expt}

\begin{figure}[h]
\centering
\includegraphics[width=8.5cm]{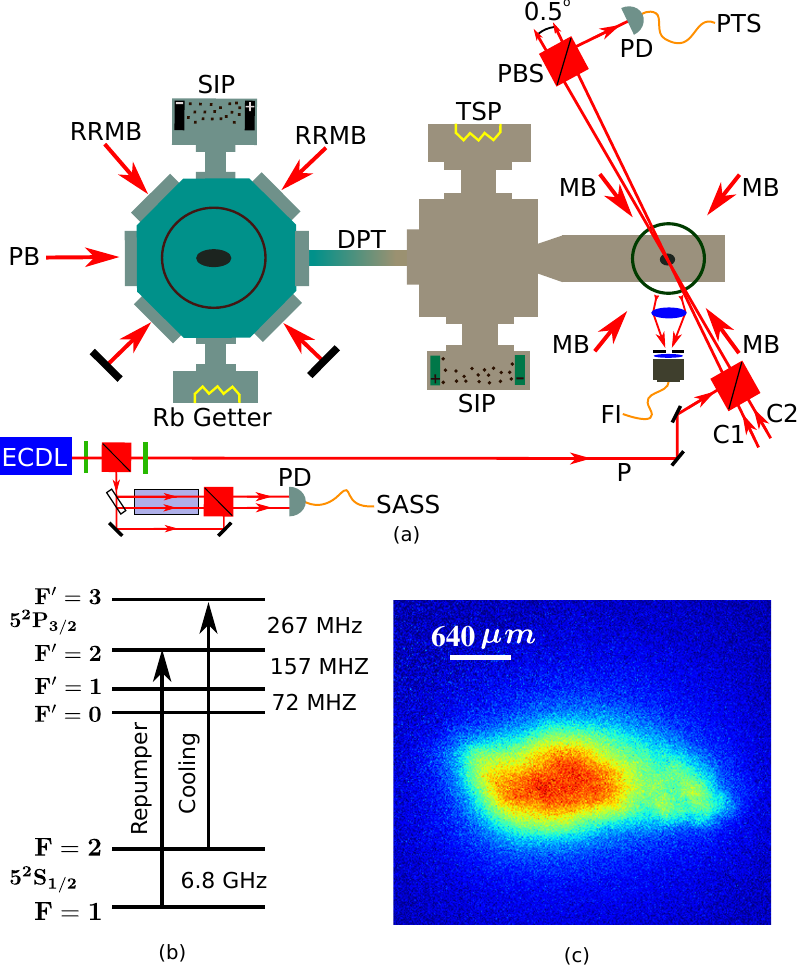}
\caption{ (color online) (a) The illustration of experimental setup. SIP: sputter-ion pump, RRMB: retro-reflected MOT beams, PB: push beam, DPT:differential pumping tube, TSP: titanium sublimation pump, MB: MOT beams, PBS: polarizing beam splitter, FI: fluorescence imaging, ECDL: extended cavity diode laser, PD: photo diode, SASS: saturated absorption spectroscopy signal, C1, C2: coupling beams, P: probe beam, PTS: probe transmission signal. (b) Energy level diagram of $^{87}Rb$ showing transition of cooling and re-pumping beams. (c) Fluorescence image of cold-atom cloud. }
\label{Fig:setup}
\end{figure}

The experiments have been performed on a double magneto-optical trap setup, the schematic of which has been illustrated in Fig. \ref{Fig:setup}(a). The vacuum chamber comprises of a stainless steel chamber (vapor chamber) kept at a pressure $\sim\ 5 \times 10^{-8}$ Torr and a quartz glass cell (ultra high vacuum (UHV) cell) kept at a pressure $\sim\ 5 \times 10^{-11}$ Torr. The two parts of the chamber were connected through a differential pumping tube with varying inner diameter to maintain the pressure gradient. Two  sputter ion pumps on both parts of the vacuum chamber along with a nonevaporable getter pump and a Titanium sublimation pump on the ultrahigh vacuum cell facilitated the required vacuum conditions. The vapor of Rb atoms are injected in vapor chamber using a Rb getter connected with the vapor chamber.  

Two magneto-optical traps (MOTs) are prepared in the setup, in which first MOT is prepared in vapor chamber (called VC- MOT) and second MOT is prepared in UHV glass cell (called as UHV-MOT). The MOTs were realised using a set of frequency locked laser beams along with a magnetic field gradient generated by pairs of anti-Helmholtz coils. For both the MOTs, the cooling beams were derived from an external cavity diode laser (Toptica, DL 100, Germany) feeding to an amplifier (TABoosta, Toptica, Germany) providing a total output power $\sim 550$ mW. The frequency of the cooling beams were locked at $\sim 15$ MHz red detuned from cooling transition frequency $5 ^{2}S_{1/2}\ F=2\ \rightarrow 5 ^{2}P_{3/2}\ F'=3 $ of $^{87}Rb$ atom (Fig. \ref{Fig:setup} (b)). Similarly, another external cavity diode laser (Toptica, DL 100, Germany) with a different amplifier (TABoosta, Toptica, Germany) gave total output power of $\sim 260$ mW to drive the repumping transition $5 ^{2}S_{1/2}\ F=1\ \rightarrow 5 ^{2}P_{3/2}\ F'=2 $ (Fig. \ref{Fig:setup} (b)). In order to realise the VC-MOT, three laser beams were applied in retro reflecting configuration with a 9 mW power in each beam (6 mW power to drive the cooling transition and 3 mW power to drive the repumping transition). The dipole force exerted by another beam (aptly named `push beam') which was 1 GHz red detuned from the cooling transition was used to extract atoms from the VC-MOT. The push beam transferred the atoms from the VC-MOT to the UHV-MOT where six independent laser beams having $1/e^2$-radius of 6 mm and 7 mW of power (5 mW power to drive the cooling transition and 2 mW power to drive the repumping transition) were directed from six directions. The magnetic field gradients in VC-MOT and UHV-MOT were $\sim\ 10\ G\ cm^{-1}$ and $\sim\ 15\ G\ cm^{-1}$ respectively. The fluoresence images of the atom cloud trapped in the UHV-MOT, shown in Fig. \ref{Fig:setup} (c), were obtained using a CCD camera (Pixelfly model USB) alongwith an appropriate 2f-imaging optics. The counts in the fluorescence image of the atom cloud were used to estimate number of atoms trapped in the UHV-MOT. The typical number of atoms in the UHV-MOT was $\sim 5 \times 10^7$ at temperature of $\sim 300\ \mu\ K$.  

The probe beam for the EIT experiments was derived from another external cavity diode laser (Toptica, DL 100, Germany) having laser linewidth less than 1 MHz and $1/e^2$-radius $\sim 3$ mm. The beam was aligned to traverse through the UHV-MOT as shown in Fig. \ref{Fig:setup} (a). In order to obtain the drive beams for the EIT experiments, a part of the laser amplifier output that drives the repumping transition in both the MOTs was extracted and passed through an accousto optic modulator (AOM) in double-pass configuration. The frequency of the output beam corresponds to the transition $5^2 S_{1/2} F=1\rightarrow 5^2 P_{3/2} F'=2$. This output beam was split using a half-waveplate and a polarizing beam splitter. One part, hereafter denoted as drive beam `$C_1$', was copropagated with the probe beam through the UHV-MOT. The other part was again passed though another AOM in double pass configuration to shift its frequency corresponding to the  $5^2 S_{1/2} F=1\rightarrow 5^2 P_{3/2} F'=1$ transition frequency. This beam, denoted as drive beam `$C_2$', was also passed through the UHV-MOT at an angle of $\sim$ 0.5 degree with the probe beam. The polarization of the probe beam was linear but orthogonal to the polarization of two drive beams (`$C_1$' and `$C_2$'). The probe beam was separated from the drive beams using a half-waveplate and a polarizing beam splitter after UHV-MOT cloud, and collected on a photodiode. The output of this photodiode was then recorded using a digital oscilloscope. The fixed frequencies of two drive beams (`$C_1$' and `$C_2$') and scan of probe beam (P) frequency forms two different N-systems in the $D_2$ line transition of $^{87}Rb$ atom simultaneously which is shown in Fig. \ref{Fig:Nsys}. The description of these N-systems is given in section \ref{sec:numsim} and \ref{sec:result}.

\section{Density Matrix Formalism}\label{sec:numsim}

\begin{figure}[t]
\centering
\includegraphics[width=8.5cm]{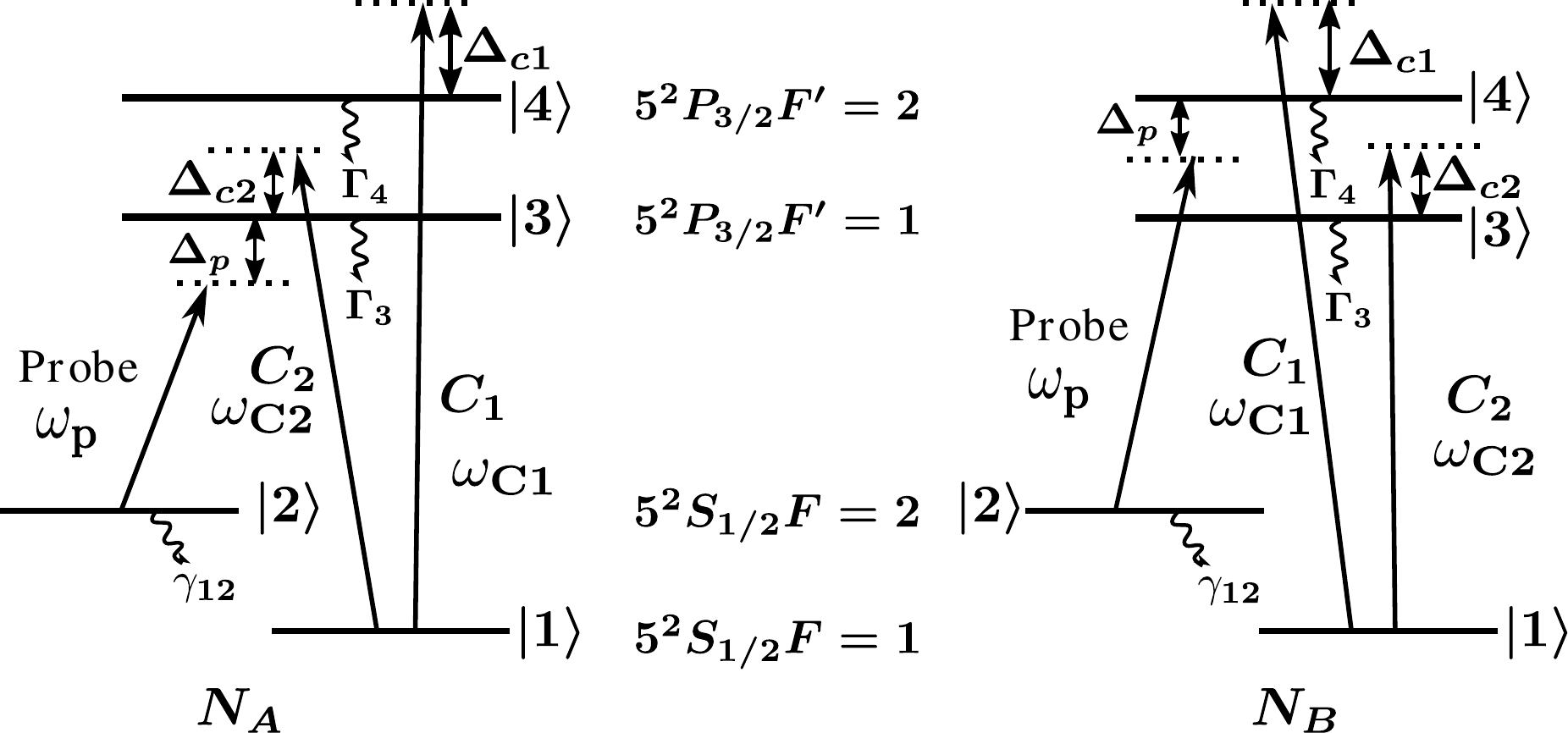}
\caption{ The energy level diagram of $^{87}Rb$ atom showing two possible N-systems $N_A$ and $N_B$. }
\label{Fig:Nsys}
\end{figure}

In order to understand the experimental observations, both the N-type systems have been studied using a density matrix formalism. The two N systems are shown in Fig. \ref{Fig:Nsys}. The detuning of the beams are defined as $\Delta_p= \omega_{32}-\omega_p$ for system $N_A$, $\Delta_p= \omega_{42}-\omega_p$ for system $N_B$, $\Delta_{c1}= \omega_{41}-\omega_{c1}$ and $\Delta_{c2}= \omega_{31}-\omega_{c2}$ for both the systems. Here, $\omega_p$, $\omega_{c1}$, and $\omega_{c2}$ are frequencies of probe beam and drive beams `$C_1$' and `$C_2$'. The density matrix formalism involves solving multiple sets of partial differential equation originated from Liouville equation. The Liouville equation in terms of the total Hamiltonian $H$ and density matrix $\rho$ is,

\begin{equation}\label{eq:liouville}
\frac{\partial\rho}{\partial t}=-\frac{i}{\hbar}\left[H,\rho\right]+\Gamma \rho.
\end{equation}
The first term in right-hand side of Eq. \ref{eq:liouville} describes the unitary evolution and second term describes the decay of the atomic system. Under the rotating wave approximation the complete density matrix evolution equations for  atomic system $N_B$ can be written as,

\begin{align*}\label{eq:L}
\frac{\partial\rho_{11}}{\partial t} &=-\frac{i}{2} \Omega_{c2} (\rho_{31}-\rho_{13}) \\& - \frac{i}{2} \Omega_{c1} (\rho_{41}-\rho_{14})+\Gamma_3 \rho_{33}+\frac{\Gamma_4}{2} \rho_{44}\nonumber\\
\frac{\partial\rho_{22}}{\partial t}&=-\frac{i}{2} \Omega_p (\rho_{42}-\rho_{24})+\Gamma_4 \rho_{44}\nonumber\\
\frac{\partial\rho_{33}}{\partial t}&=-\frac{i}{2} \Omega_{c2} (\rho_{13}-\rho_{31})-\Gamma_3\rho_{33}\nonumber\\
\frac{\partial\rho_{44}}{\partial t}&=-\frac{i}{2} \Omega_p (\rho_{24}-\rho_{42})+\frac{i}{2} \Omega_{c1} (\rho_{41}-\rho_{14})-\Gamma_4 \rho_{44}\nonumber\\
\frac{\partial\rho_{12}}{\partial t}&=\left(-i(\Delta_p-\Delta_{c1})-\gamma_{12}\right) \rho_{12}-\frac{i}{2} \Omega_{c2}\rho_{32}\\&+\frac{i}{2} \Omega_{p}\rho_{14}-\frac{i}{2} \Omega_{c1}\rho_{42}\nonumber\\
\frac{\partial\rho_{13}}{\partial t}&=\left(i\Delta_{c2}-\frac{\Gamma_3}{2} \right) \rho_{13}-\frac{i}{2} \Omega_{c1}\rho_{43}\\&+\frac{i}{2} \Omega_{c2}(\rho_{11}-\rho_{33})\nonumber\\
\frac{\partial\rho_{14}}{\partial t}&=\left(i\Delta_{c1}-\frac{\Gamma_4}{2}\right) \rho_{14}+\frac{i}{2} \Omega_{p}\rho_{12}\\&-\frac{i}{2} \Omega_{c2}\rho_{34}+\frac{i}{2} \Omega_{c1}(\rho_{11}-\rho_{44})\nonumber\\
\frac{\partial\rho_{23}}{\partial t}&=\left(i(\Delta_p+\Delta_{c2}-\Delta_{c1})-\frac{\Gamma_3}{2}\right) \rho_{23}\\&-\frac{i}{2} \Omega_{p}\rho_{43}+\frac{i}{2} \Omega_{c2}\rho_{21}\nonumber\\
\frac{\partial\rho_{24}}{\partial t}&=\left(i\Delta_p-\frac{\Gamma_4}{2}\right) \rho_{24}\\&+\frac{i}{2} \Omega_{p}(\rho_{22}-\rho_{44})+\frac{i}{2} \Omega_{c1}\rho_{21}\nonumber\\
\frac{\partial\rho_{34}}{\partial t}&=\left(i(\Delta_{c1}-\Delta_{c2})-\frac{\Gamma_3 + \Gamma_4}{2}\right) \rho_{34}\\&+\frac{i}{2} \Omega_{p}\rho_{32}+\frac{i}{2} \Omega_{c1}\rho_{31}-\frac{i}{2} \Omega_{c2}\rho_{14}\nonumber\\
\end{align*}

with, 
\begin{align}
\frac{\partial\rho_{ij}}{\partial t}&=\frac{\partial\rho_{ji}^*}{\partial t}.
\end{align}

Similarly, set of density matrix equations for atomic system $N_A$ can also be derived. In order to solve the above set of equations in the steady state, the matrix method is used. In this method, the solution for density matrix elements is obtained from the equation,

\begin{equation}\label{eqn:L2}
\dot{\rho}=\mathcal{L}\rho=0.
\end{equation}

The steady state value of the density matrix $\rho$ is an eigenvector of the Liouvillian super-operator ($\mathcal{L}$) corresponding to the `zero' eigenvalue. 

\section{Results and Discussion}
\label{sec:result}

\begin{table*}[ht]
\caption{\label{tab:transitions} Transitions involved in various systems}
\centering 
\begin{tabular*}{\textwidth}{c @{\extracolsep{\fill}} cccc}
\hline \hline 
system & coupling & probe & control  \\ [0.5ex]
\hline 
$\Lambda_A$ & $5^2 S_{1/2} F=1\rightarrow 5^2 P_{3/2} F'=1$ & $5^2 S_{1/2} F=2\rightarrow 5^2 P_{3/2} F'=1$ & -  \\  
$\Lambda_B$ & $5^2 S_{1/2} F=1\rightarrow 5^2 P_{3/2} F'=2$ & $5^2 S_{1/2} F=2\rightarrow 5^2 P_{3/2} F'=2$ & -  \\  
$N_A$ & $5^2 S_{1/2} F=1\rightarrow 5^2 P_{3/2} F'=1$ & $5^2 S_{1/2} F=2\rightarrow 5^2 P_{3/2} F'=1$ & $5^2 S_{1/2} F=1\rightarrow 5^2 P_{3/2} F'=2$  \\  
$N_B$ & $5^2 S_{1/2} F=1\rightarrow 5^2 P_{3/2} F'=2$ & $5^2 S_{1/2} F=2\rightarrow 5^2 P_{3/2} F'=2$ & $5^2 S_{1/2} F=1\rightarrow 5^2 P_{3/2} F'=1$  \\ [0.5ex]
\hline     
\end{tabular*}
\end{table*}

The N-type atomic system is an extension of the basic $\Lambda$-system where an additional laser beam, called as the control beam, couples one of the ground state with another excited state. The studied N-type systems are shown in Fig. \ref{Fig:Nsys}. The first $\Lambda$-system denoted as $\Lambda_A$, formed using the transitions $5 ^{2}S_{1/2}\ F=1\ \rightarrow 5 ^{2}P_{3/2}\ F'=1\ \leftarrow\ 5 ^{2}S_{1/2}\ F=2$, was converted to an N-type atomic system $N_A$ by adding another drive beam `$C_1$' resonant to transition  $5 ^{2}S_{1/2}\ F=1\ \rightarrow 5 ^{2}P_{3/2}\ F'=2$. Likewise, the atomic system $\Lambda_B$ formed using the transitions $5 ^{2}S_{1/2}\ F=1\ \rightarrow 5 ^{2}P_{3/2}\ F'=2\ \leftarrow\ 5 ^{2}S_{1/2}\ F=2$, was converted to another N-type atomic system $N_B$ in presence of drive beam `$C_2$' resonant to transition $5 ^{2}S_{1/2}\ F=1\ \rightarrow 5 ^{2}P_{3/2}\ F'=1$. The coupling, probe and control beams frequencies for different systems ($\Lambda_A$, $\Lambda_B$, $N_A$ and $N_B$) are shown in table \ref{tab:transitions}. We note that the drive beam `$C_1$' acts as a control beam for system $N_A$, whereas drive beam `$C_2$' acts as a control beam for system $N_B$.

As the strength of a transition is governed by the Clebsch-Gordan (CG) coefficient connecting two states, the probe  transition strength at a given probe beam power is higher in atomic system $N_B$ than that in system $N_A$. However, the CG coefficients of both the driving beams `$C_1$' and `$C_2$' are same giving same strength for equal powers in these beams. In the experiments, probe power was kept fixed at $\sim$ 30 $\mu$ W. In all the studies, the transmitted probe signal for both the atomic systems (A and B) was captured in a single scan of probe detuning using a digital storage oscilloscope, but the results of two systems are presented in separate plots. The experimental data shown in the Fig.s have been processed for better visibility using a convolution algorithm.

\begin{figure}[t]
\centering
\includegraphics[width=8.5cm]{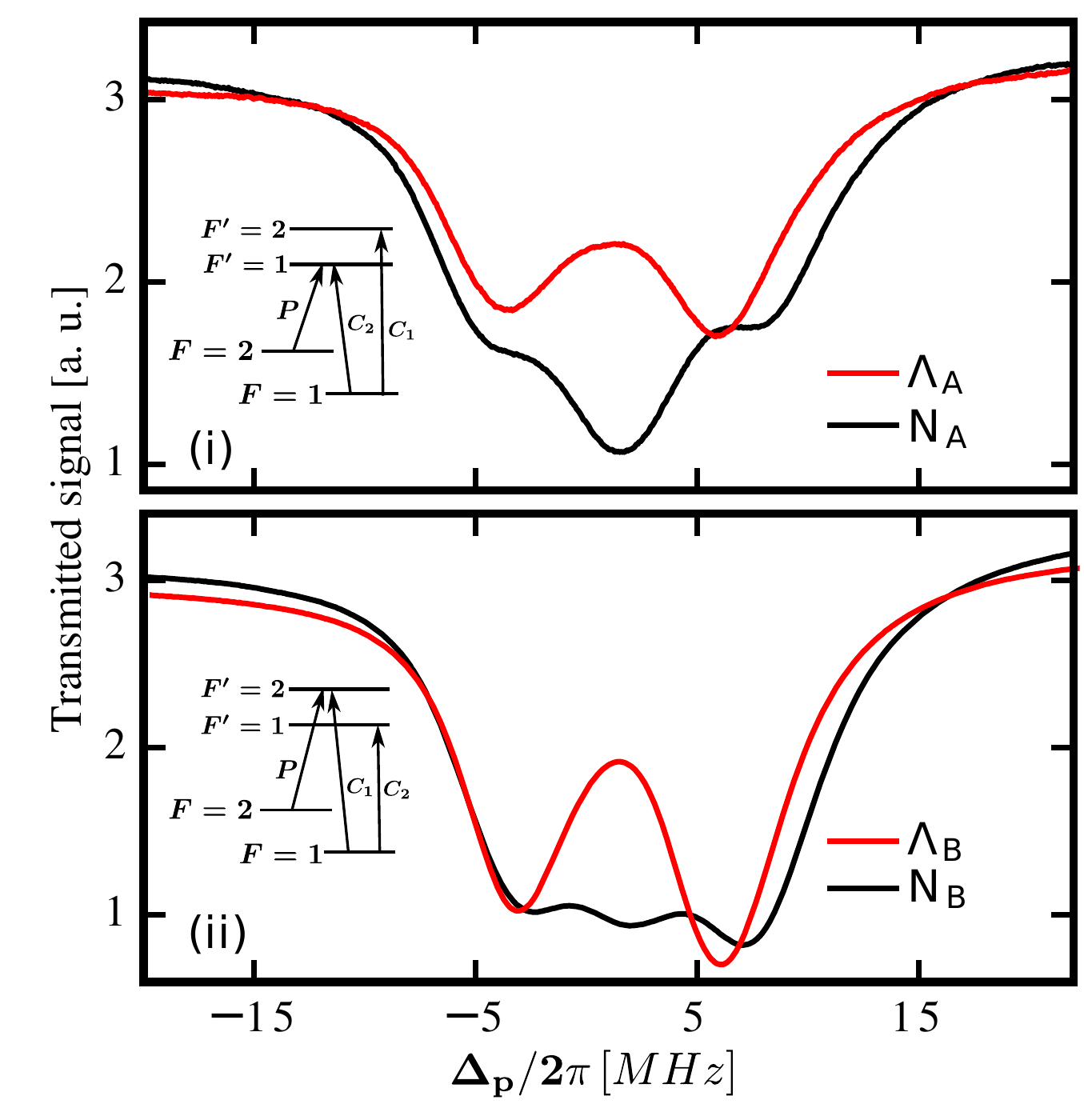}
\caption{ (Color online) The transmitted probe signal as a function of probe beam detuning for $\Lambda$- (shown by red curves) and N- (shown by black curves) systems. Here plots (i) and (ii) correspond to systems (A) and (B) respectively, with both driving beams '$C_1$' and `$C_2$' are of power 12 mW and probe beam of power $\sim$ 30 $\mu$W.}
\label{Fig:comp}
\end{figure}

\begin{figure}[t]
\centering
\includegraphics[width=8.5cm]{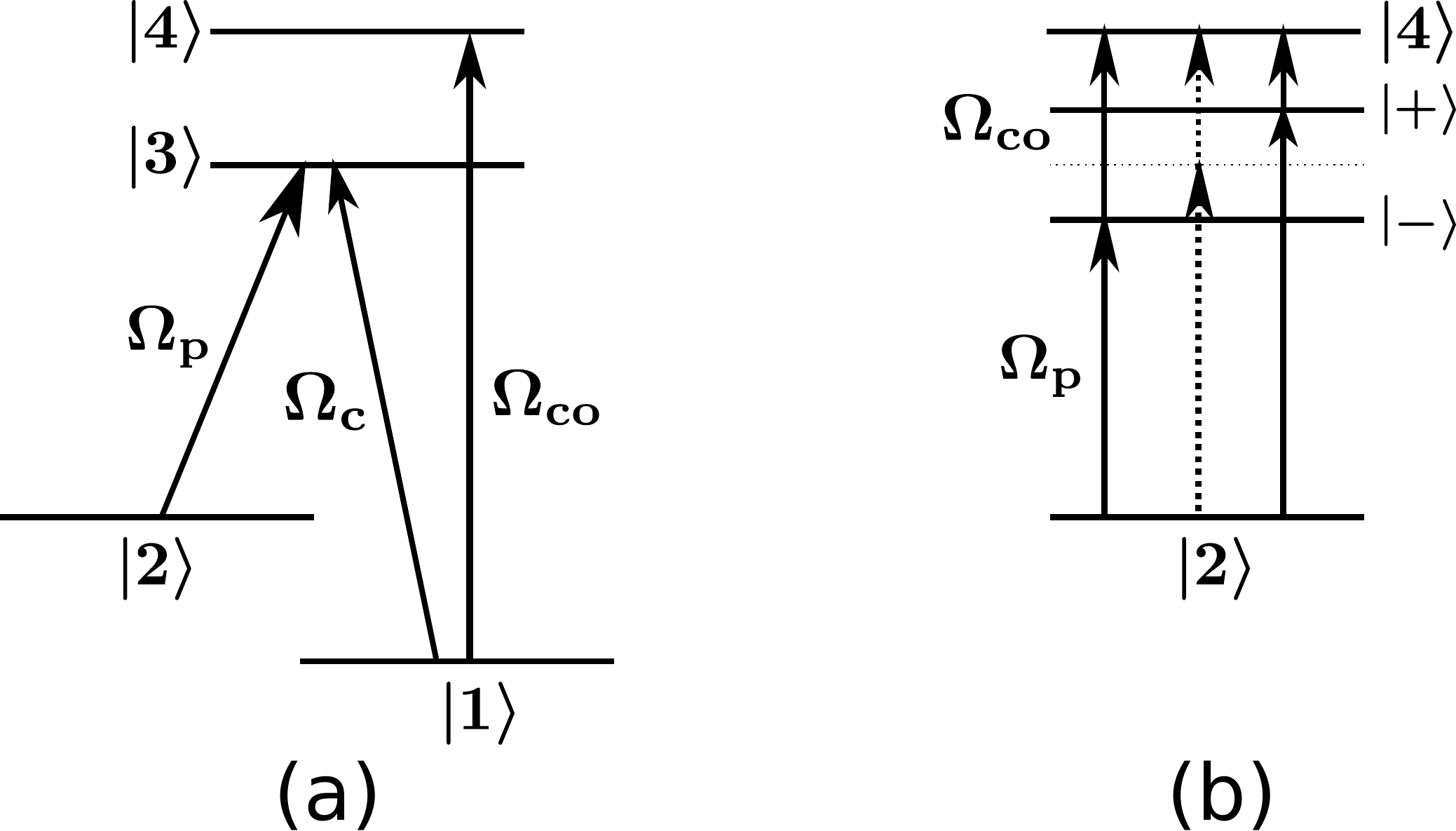}
\caption{ (a) The bare system of four-level N-system with coupling field strength $\Omega_c$, control field strength $\Omega_{co}$ and probe field strength $\Omega_p$. (b) The formation of dressed-state due to coupling field and two possible paths for transition from ground state $|2\rangle$ to excited state $|4\rangle$ shown by solid arrows. The dotted arrows show enhanced two-photon absorption due to constructive quantum interference.}
\label{Fig:dress}
\end{figure}

The transmitted probe spectrum for both the $\Lambda$-systems and corresponding N-systems were measured. In these measurements, both the driving beam frequencies were kept at resonance with equal power (12 mW) in both the beams. The observed results for system (A) and (B) are shown in Fig. \ref{Fig:comp} (i) and (ii) respectively. In the absence of the control beams, `$C_1$' for system (A) and `$C_2$' for system (B), the transmitted probe signals have shown the expected EIT features of the $\Lambda$-systems. As has been reported earlier \cite{Charu:65:2018}, the EIT signal in system $\Lambda_B$ was observed to be stronger than that in system $\Lambda_A$. In the presence of control beam, both the systems ($N_A$ and $N_B$) have shown a transmission dip at the center of the existing EIT peak, resulting in splitting of the single EIT peak into two separate EIT peaks \textit{i.e.} dual EIT, shown by black curves in Fig. \ref{Fig:comp} (i) and (ii). The emergence of central transmission dip in the presence of control beam is due to three-photon process at the line center \textit{i.e.} $|2\rangle \rightarrow |3\rangle \rightarrow |1\rangle \rightarrow |4\rangle$ for system $N_A$ and $|2\rangle \rightarrow |4\rangle \rightarrow |1\rangle \rightarrow |3\rangle$ for system $N_B$. In the dressed picture, this central transmission dip is a result of enhanced two-photon absorption (TPA) occurring due to constructive quantum interference between paths $|2\rangle \rightarrow |+\rangle \rightarrow |4\rangle$ and $|2\rangle \rightarrow |-\rangle \rightarrow |4\rangle$. The states $|+\rangle$ and $|-\rangle$ are dressed states created by coupling beam (`$C_2$' for system $N_A$ and `$C_1$' for system $N_B$) \cite{Mulchan:2000, Yan:26:2001}. The side transmission dips in Fig. \ref{Fig:comp} (black curves) are Autler Townes (AT) absorption, which are due to single photon absorption from state $|2\rangle$ to the dressed states $|+\rangle$ and $|-\rangle$. The pictorial representation of enhanced TPA due to constructive quantum interference in N-system is shown in Fig. \ref{Fig:dress}. In this article, the side transmission dips are referred as AT transmission dips and central dip is referred as TPA transmission dip. Both the N-systems have shown same spectral feature except that system $N_A$ exhibits greater TPA transmission dip than the AT transmission dips. The enhanced absorption at line centre due to the presence of additional control field opens the path for the system to be useful in optical switching devices.

\begin{figure}[t]
\centering
\includegraphics[width=8.5cm]{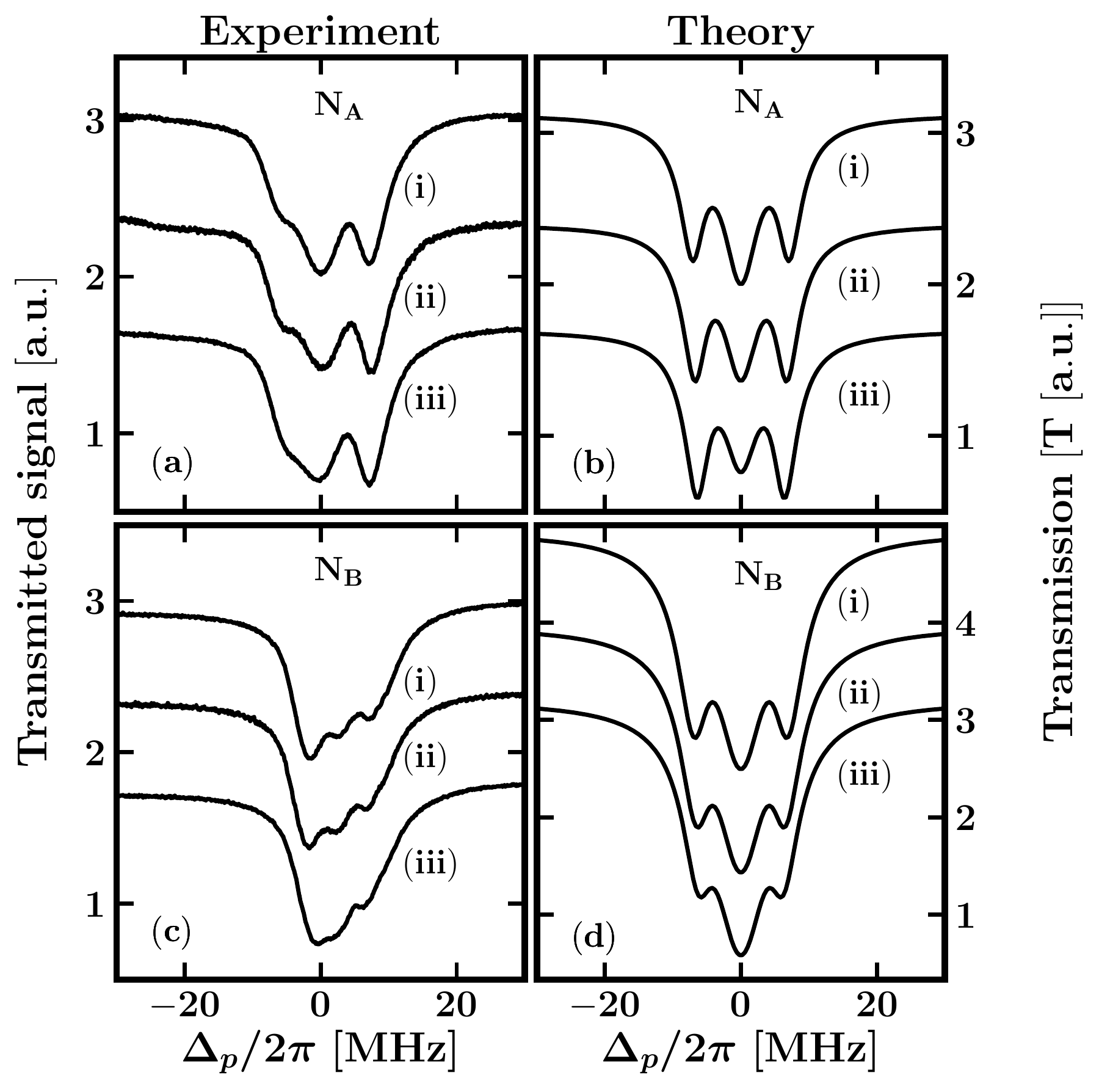}
\caption{ (a) and (c) show the transmitted probe beam signal as a function of the probe beam detuning $\Delta_p$ in systems $N_A$ and $N_B$ for different values of $P_{c1}$ and fixed value of $P_{c2}=$ 12 mW. The power values in `$C_1$' beam $P_{c1}$ are : \textit{(i)} 12 mW, \textit{(ii)} 10 mW, and \textit{(iii)} 8 mW. Plots (b) and (d) show the calculated transmission as a function of the probe beam detuning $\Delta_p$ with $\Omega_{c1}$ as: \textit{(i)} $2 \pi \times$ 10 MHz, \textit{(ii)} $2 \pi \times$ 9 MHz, and \textit{(iii)} $2 \pi \times$ 8 MHz, for a fixed $\Omega_{c2}= 2 \pi \times$ 10.0 MHz, for both the systems $N_A$ and $N_B$. The other parameters used in the numerical simulations for plot (b) are $\Omega_p= 2 \pi \times$ 0.6 MHz,  $\Delta_{c1} = \Delta_{c2}= 2 \pi \times$ 0.0 MHz, and for plot (d) are $\Omega_p= 2 \pi \times$ 1.3 MHz, $\Delta_{c1} = \Delta_{c2}= 2 \pi \times$ 0 MHz. }
\label{Fig:Pc1}
\end{figure}

To investigate the atomic systems $N_A$ and $N_B$ further, the power of one of the driving beams was varied while keeping the other one fixed. The other experimental parameters were kept unchanged during these measurements. First, we varied power in drive beam `$C_1$', while keeping the power of drive beam `$C_2$' fixed at 12 mW. The observed transmitted probe spectrum is shown in Fig. \ref{Fig:Pc1} (a) and (c). For system $N_A$, with decrease in power of drive beam `$C_1$', the strength of the central transmission dip reduces (as shown in Fig. \ref{Fig:Pc1} (a)). This is because, the drive beam `$C_1$' acts as a control beam for system $N_A$, which contributes to TPA feature. Hence, the reduction in intensity of beam `$C_1$' reduces the strength of TPA feature (\textit{i.e.} central transmission dip). The numerically calculated probe transmission for system $N_A$ has also shown similar behaviour and is presented in Fig. \ref{Fig:Pc1} (b). For system $N_B$, as observable from the Fig. \ref{Fig:Pc1} (c), with the decrease in power in beam `$C_1$', the AT transmission dips get reduced with nearly unchanged strength of TPA transmission dip. Since the drive beam `$C_1$' acts as a coupling beam for system $N_B$, its effect on AT transmission dips is higher than that on the TPA  transmission dip. The numerical results for system $N_B$, obtained by solving Eq. \ref{eq:L} and \ref{eqn:L2}, are also shown in Fig. \ref{Fig:Pc1} (d) which are in qualitative agreement with the experimental observations shown in Fig. \ref{Fig:Pc1} (c). In the simulation, the field's Rabi frequencies were kept close to the experimental one and decay rates were $\Gamma_3=\Gamma_4=2 \pi \times 6\ MHz$ \textit{i.e.} decay rate of excited states in $D_2$ line of $^{87}Rb$ atom and $\gamma_{12}=2 \pi \times 1\ MHz$ for incorporating laser linewidth. We note some asymmetry in the experimentally observed results as compared to the simulation results. This may be attributed to presence of neighbouring levels in real atomic system, broadening in energy levels due to magnetic field of MOT and spatial inhomogeneity in field's Rabi frequency, which are not considered in the simulations. 

\begin{figure}[t]
\centering
\includegraphics[width=8.5cm]{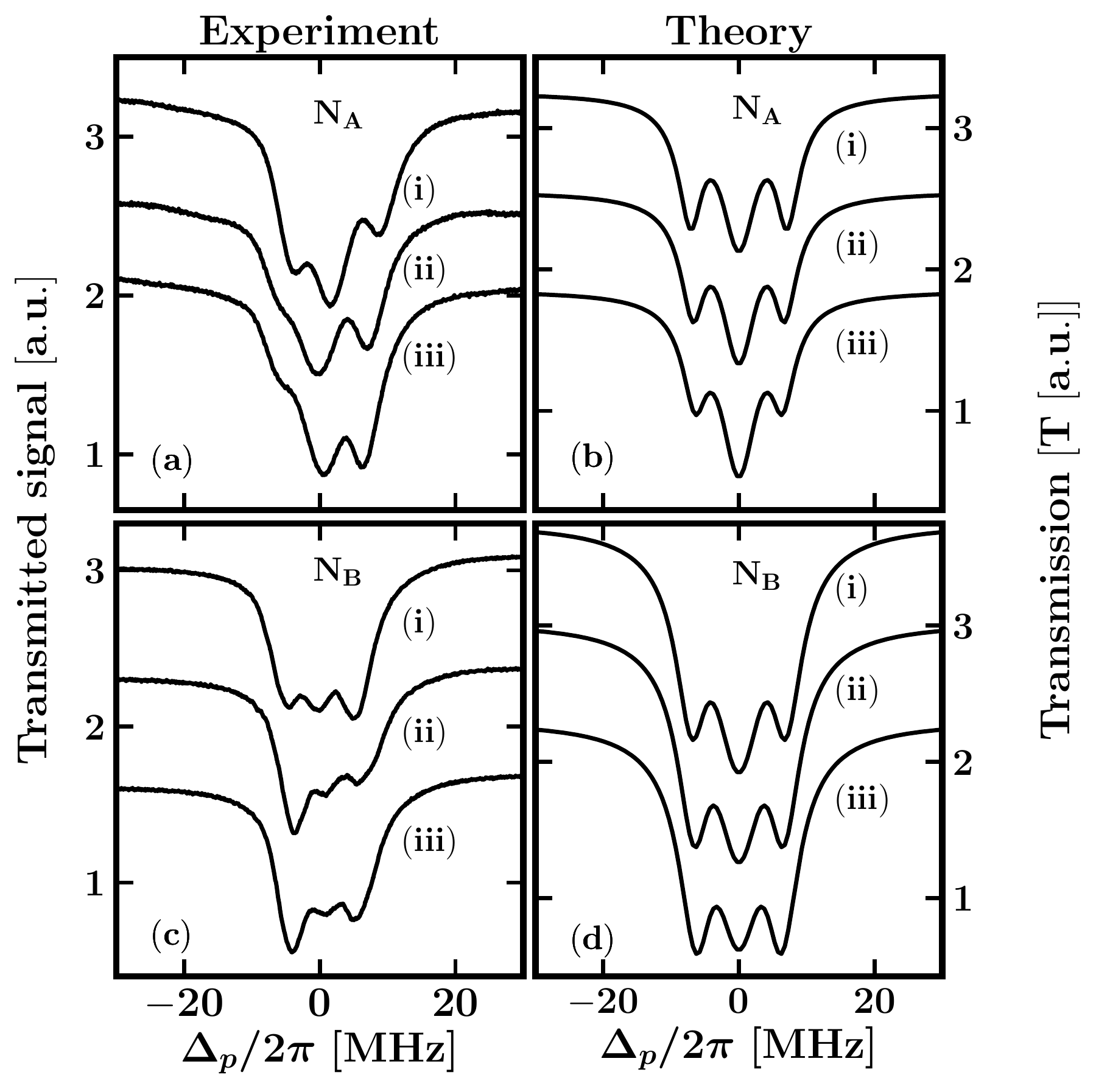}
\caption{ (a) and (c) show the transmitted probe beam signal as a function of the probe beam detuning $\Delta_p$ in systems $N_A$ and $N_B$ for different values of $P_{c2}$ and fixed value of $P_{c1}=$ 12 mW. The power values in `$C_2$' beam $P_{c2}$ are : \textit{(i)} 12 mW, \textit{(ii)} 10 mW, and \textit{(iii)} 8 mW. Plots (b) and (d) show the calculated transmission as a function of the probe beam detuning $\Delta_p$ with $\Omega_{c2}$ as: \textit{(i)} $2 \pi \times$ 10 MHz, \textit{(ii)} $2 \pi \times$ 9 MHz, and \textit{(iii)} $2 \pi \times$ 8 MHz, for a fixed $\Omega_{c1}= 2 \pi \times$ 10.0 MHz, for both the systems $N_A$ and $N_B$. The other parameters used in the numerical simulations for plot (b) are $\Omega_p= 2 \pi \times$ 0.6 MHz,  $\Delta_{c1} = \Delta_{c2}= 2 \pi \times$ 0 MHz, and for plot (d) are $\Omega_p= 2 \pi \times$ 1.3 MHz, $\Delta_{c1} = \Delta_{c2}= 2 \pi \times$ 0 MHz.}
\label{Fig:Pc2}
\end{figure}

Next, the power of driving beam `$C_2$' $P_{c2}$ was varied from 12 mW to 8 mW, while keeping $P_{c1}$ fixed at 12 mW and the corresponding results are shown in Fig. \ref{Fig:Pc2} (a) and (c). Since the driving beam `$C_2$' is the coupling field for atomic system $\Lambda_A$ and $N_A$, the reduction in strength of the AT Transmission dips is an observation consistent with the results obtained with the variation in power of driving beam `$C_1$' for system $N_B$ discussed before. For the system $N_B$, the driving beam `$C_2$' acts as a control field, therefore as expected, the decrease in the strength of the TPA transmission dip with the decrease in $P_{c2}$ in Fig. \ref{Fig:Pc2} (c) is similar to the observations presented in Fig. \ref{Fig:Pc1} (a) for variation in `$C_1$' beam power. The corresponding calculated spectra for both the systems $N_A$ and $N_B$ are shown in Fig. \ref{Fig:Pc2} (b) and (d)respectively, which have shown an agreement with the experimental observations.

\begin{figure}[t]
\centering
\includegraphics[width=8.5cm]{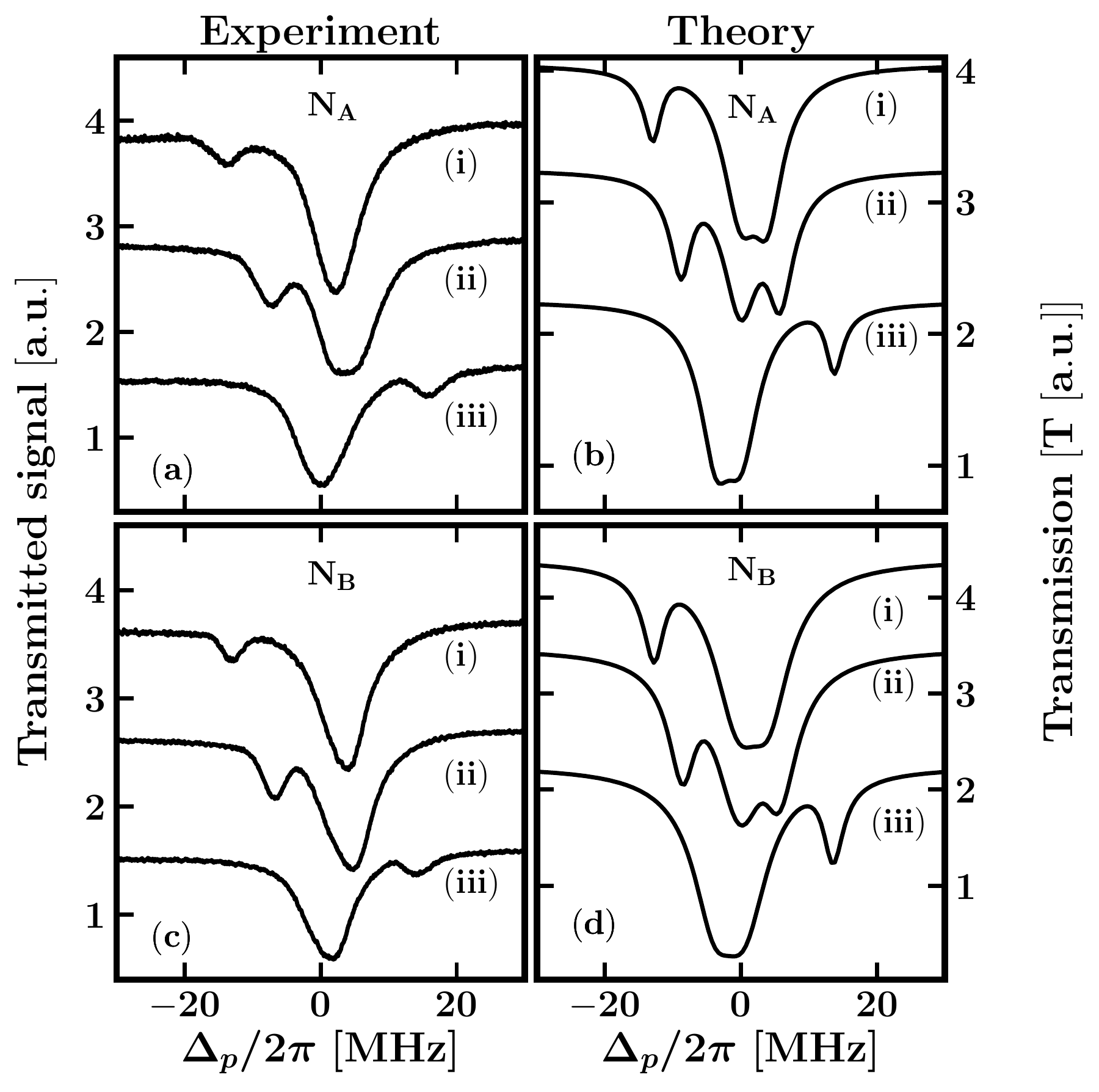}
\caption{ (a) and (c) show the measured probe transmission spectrum and (b) and (d) show the calculated probe transmission spectrum for systems $N_A$ and $N_B$, for $\Delta_{c1}$ and $\Delta_{c2}$:  \textit{(i)} $\sim$ - $2 \pi \times$ 9 MHz, \textit{(ii)} $\sim$ - $2 \pi \times$ 3 MHz, \textit{(iii)} $\sim$ $2 \pi \times$ 10 MHz. Powers of both the driving beams are 12 mW and corresponding Rabi strength in simulation was considered $\Omega_{c1} =\Omega_{c2}=2 \pi \times$ 10 MHz. The other parameters used in simulations are $\Omega_p= 2 \pi \times$ 0.6 MHz and $2 \pi \times$ 1.3 MHz for systems $N_A$ (plot (b)) and $N_B$ (plot (d)) respectively.}
\label{Fig:delC1C2}
\end{figure}

The dependence of probe transmission spectrum on detuning of both the driving beams for both the atomic systems $N_A$ and $N_B$ has also been studied. For this study, the driving beams frequency detunings were kept equal ($\Delta_{c1}=\Delta_{c2}$), but varied from -$2 \pi \times$ 9 MHz to $2 \pi \times$ 10 MHz. The power in both the beams during this measurement was $\sim$12 mW. The corresponding experimental and numerical results are shown in Fig. \ref{Fig:delC1C2}. The equal detuning of both the drive beams means equal detuning of coupling and control beam for both the N-systems. As the control beam is far detuned, it acts as a perturbation to the corresponding $\Lambda$-system and does not contribute much to the two-photon absorption process. With detuned coupling beam, the system becomes detuned $\Lambda$-system. Thus, the N-system becomes detuned $\Lambda$-system with perturbation when both the drive beams are far detuned. From the Fig. \ref{Fig:delC1C2} (a) and (c), it is observable that when control beam is far detuned (say -9 and 10 MHz) the system has shown asymmetric transmission dips which is a signature of detuned $\Lambda$ system. For low detuning value ($\sim$ -3 MHz), the presence of control beam results in broadening in transmission dips. For this value of detuning ($\Delta_{c1}=\Delta_{c2}=-3$ MHz) the simulation results (curve (ii) in Fig. \ref{Fig:delC1C2})have shown resolution between three dips which is absent in experimental results. This may be because of neglecting presence of neighbouring levels, Zeeman splitting due to magnetic field of MOT, etc in our numerical simulations. This study reveals that simultaneous detuning of both the drive beams can alter the number of transmission dips in the atomic medium.

\begin{figure}[t]
\centering
\includegraphics[width=8.5cm]{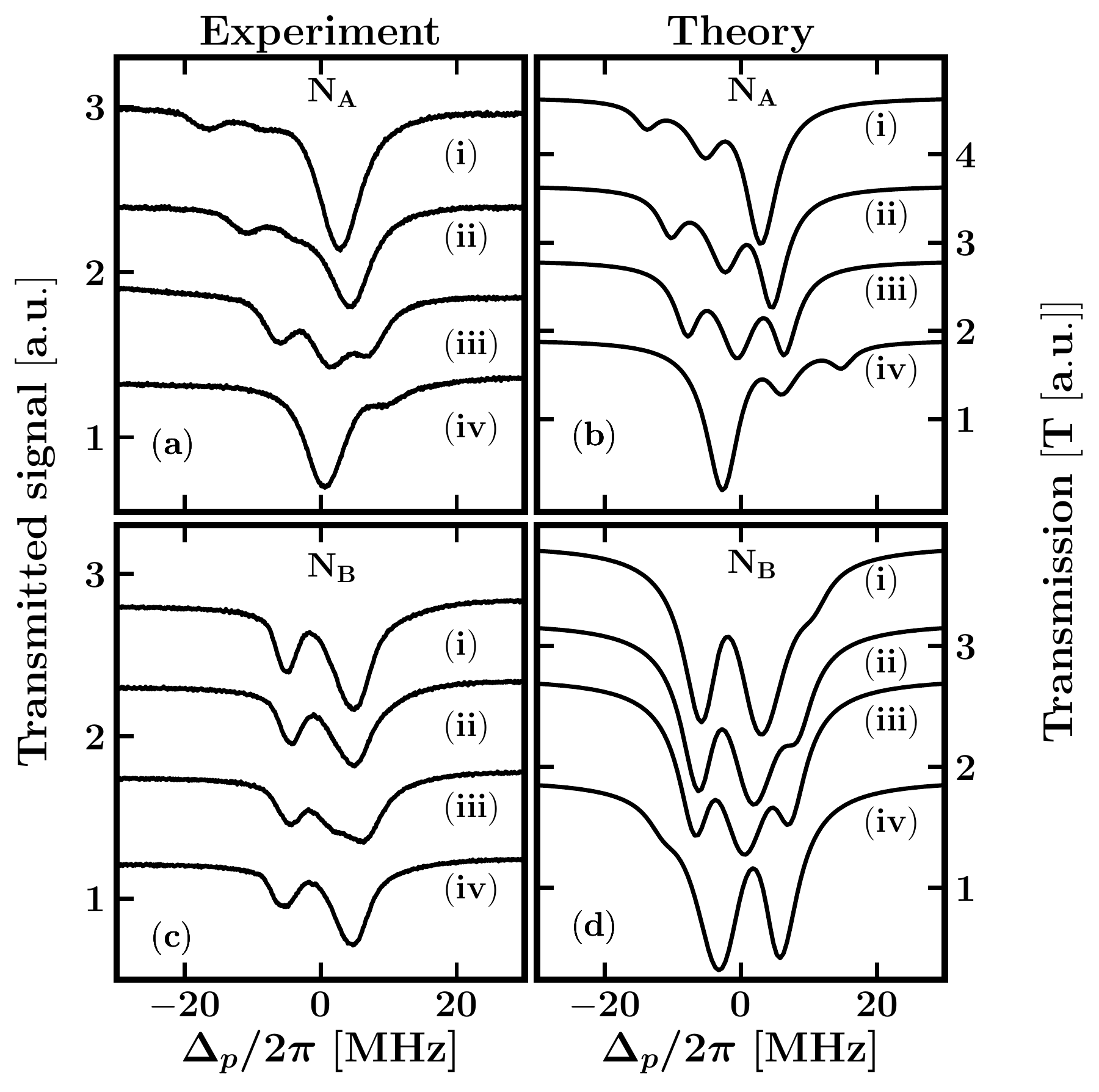}
\caption{ (a) and (c) show the transmitted probe beam signal as a function of the probe beam detuning $\Delta_p$ in systems $N_A$ and $N_B$ for different detuning values of the drive beam `$C_2$' ($\Delta_{c2}$) with $\Delta_{c1}=0$ and drive beam powers $P_{c1}=P_{c2}=$ 12 mW. Plot (b) and (d) show the calculated transmission as a function of the probe beam detuning $\Delta_p$ for different detuning values of the coupling field $\Delta_{c2}$ with $\Delta_{c1}=0$ and field strengths $\Omega_{c1}=\Omega_{c2}=2 \pi \times$ 10 MHz. In all plots, the corresponding detuning values $\Delta_{c2}$ are \textit{(i)} $\sim$ - $2 \pi \times$ 8 MHz, \textit{(ii)} $\sim$ - $2 \pi \times$ 4 MHz, \textit{(iii)} $\sim$ $2 \pi \times$ 1 MHz, \textit{(iv)} $\sim 2 \pi \times$ 9 MHz. The other parameters used in simulations are $\Omega_p= 2 \pi \times$ 0.6 MHz and $2 \pi \times$ 1.3 MHz for systems $N_A$ and $N_B$ respectively.}
\label{Fig:delC2}
\end{figure}

The effect of detuning of the coupling and control beam independently on the probe transmission spectrum has also been studied by detuning only the drive beam `$C_2$' while keeping the other drive beam `$C_1$' fixed at resonance. In this experimental scheme, the coupling beam detuning effects can be noted from the spectral features of the system $N_A$ while the control beam detuning effect can be observed from the system $N_B$. For system $N_A$, the experimental as well as numerical results are shown in Fig. \ref{Fig:delC2} (a) and (b). For the far red detuned coupling field, the two transmission dips appear at far red detuned positions and one transmission dip appears at resonance in the probe transmission spectrum (curve (\textit{i}) in Fig. \ref{Fig:delC2} (a) and (b)). For blue detuned coupling field `$C_2$', similar spectral features appear in probe transmission spectrum with positions of dips swapped towards blue side (curve (\textit{iv}) in Fig. \ref{Fig:delC2} (a) and (b)). For near resonant case (curve (\textit{iii}) in Fig. \ref{Fig:delC2} (a) and (b) ), the spectral features recovered to three transmission dips \textit{i.e.} dual EIT. The detuned coupling field shift the dressed states $|+\rangle$ and $|-\rangle$ (refer Fig. \ref{Fig:dress}) created by coupling beam which eventually shifts the AT transmission dips. The line center of two dressed states $|+\rangle$ and $|-\rangle$ also shifts, resulting in frequency shift of TPA transmission dip. For system $N_B$, the variation in detuning of beam `$C_2$' will reveal the effect of control beam detuning variation on the probe transmission spectrum. This effect has been presented in Fig. \ref{Fig:delC2} (c) and (d). The far detuned control field has resulted in a single EIT peak at resonance (shown by curve (\textit{i}),(\textit{ii}) and (\textit{iv})). This signifies that far detuned control field has negligible effect on the spectral feature of corresponding $\Lambda$-system (i.e. $\Lambda_B$). In summary, the above studies suggest that the detuning of drive beams control the spectral features of the probe transmission considerably. The single EIT peak can be made to dual EIT peak, and vice-versa, by adjusting the detuning of the control fields.

\section{CONCLUSION}
\label{sec:concl}

The dual EIT peaks have been observed in two different N-systems in $D_2$ line transition of cold $^{87}Rb$ atoms trapped in a magneto-optical trap. The dual EIT feature is due to appearance of two-photon absorption (TPA) transmission dip in the presence of the control beam in the N-system which modifies the Autler-Townes spectrum of the given $\Lambda$-system.  Each of the two systems has shown dual EIT peaks in the transmission of probe whose dependence on coupling and control beams strength and detuning has been investigated. The results have shown that the control and coupling beams of N-system govern the central and side transmission dips respectively. Using simultaneously far detuned coupling and control beams, the dual EIT peaks diminished and two transmission dips with unequal magnitudes appeared for each N-system. The spectral positions of these transmission dips in probe spectrum depend on the sign and magnitude of coupling and control beams detuning. By varying the detuning of one drive beam, while keeping other drive beam at resonance, one N-system (for which the varying beam acted as a coupling beam) has shown three transmission dips while another N-system (for which the varying beam acted as a control beam) has shown a large single EIT peak like in $\Lambda$-system. The numerical model has shown a qualitative agreement to the experimentally observed results. The two dual EIT peaks (\textit{i.e.} four EIT peaks), for the case of resonant drive beams with equal power, provide an opportunity to explore this system further for application in multi-channel optical communication, slow light propagation, etc.

\vspace{0.5 cm}

\section{ACKNOWLEDGMENTS}
Charu Mishra is grateful for financial support from RRCAT, Indore under HBNI, Mumbai program.

\section*{References}


\end{document}